\documentclass[%
reprint,
superscriptaddress,
 amsmath,amssymb,
 aps,
prl,
]{revtex4-2}

\usepackage{ulem}

\usepackage{graphicx}
\usepackage{bm}
\newcommand{\ldna}{$\lambda$DNA~}
\usepackage[dvipsnames]{xcolor}

\newcommand{\rev}[1]{\textcolor{black}{#1}}
\newcommand{\rrev}[1]{\textcolor{black}{#1}}
\begin{document}

\title{Fluidification of entanglements by a DNA bending protein} 

%%%

%%%
\author{Yair A.~G. Fosado}
\thanks{joint first author}
\affiliation{School of Physics and Astronomy, University of Edinburgh, Peter Guthrie Tait Road, Edinburgh, EH9 3FD, UK}
\author{Jamieson Howard}
\thanks{joint first author}
\affiliation{School of Physics, Engineering and Technology, University of York, York, YO10 5DD, UK}
\author{Simon Weir}
\affiliation{School of Physics and Astronomy, University of Edinburgh, Peter Guthrie Tait Road, Edinburgh, EH9 3FD, UK}
\author{Agnes Noy}
\affiliation{School of Physics, Engineering and Technology, University of York, York, YO10 5DD, UK}
\author{Mark C Leake}
\thanks{co-corresponding author, mark.leake@york.ac.uk}
\affiliation{School of Physics, Engineering and Technology, University of York, York, YO10 5DD, UK}
\affiliation{Department of Biology, University of York, York, YO10 5DD, UK}
\author{Davide Michieletto}
\thanks{corresponding author, davide.michieletto@ed.ac.uk}
\affiliation{School of Physics and Astronomy, University of Edinburgh, Peter Guthrie Tait Road, Edinburgh, EH9 3FD, UK}
\affiliation{MRC Human Genetics Unit, Institute of Genetics and Cancer, University of Edinburgh, Edinburgh EH4 2XU, UK}

\begin{abstract}
\textbf{In spite of the nanoscale and single-molecule insights into nucleoid associated proteins (NAPs), their role in modulating the mesoscale viscoelasticity of entangled \rrev{DNA} has been overlooked so far. By combining microrheology and molecular dynamics simulation we find that the abundant NAP ``Integration Host Factor'' (IHF) lowers the viscosity of entangled $\lambda$DNA 20-fold at physiological concentrations and stoichiometries. Our results suggest that IHF may play a previously unappreciated role in resolving DNA entanglements and in turn may be acting as a ``genomic fluidiser'' for bacterial genomes.} \\
\end{abstract}

\maketitle

\vspace{-1.5 cm}
Prokaryotic and eukaryotic genomes carry out complex biological tasks which would be impossible if randomly folded~\cite{jerkovic2021,new2,Verma2019a,Japaridze2020,Gogou2021}. In bacteria, nucleoid-associated proteins (NAPs)~\cite{Verma2019a} play an important role in folding the genome~\cite{Badrinarayanan2015,Verma2019a,Dame2020,Wu2019}. Single-molecule techniques have shed light into how certain NAPs bind, bend, kink, coat or stiffen \rev{short DNA molecules in dilute conditions} ~\cite{Ali2001,Dame2006a,Liu2010c,Liang2017a,Dame2020,Yoshua2021,Japaridze2021}. However, we \rev{have little to no evidence} on what is their impact on entangled and crowded DNA~\cite{Badrinarayanan2015}. For instance, while DNA segregation is impaired when NAPs are removed from the cell~\cite{Gogou2021,Wu2019}, the NAP-mediated mechanisms through which this segregation is achieved remain to be determined. Here, we focus on the Integration Host Factor (IHF), an abundant NAP, present at about 6,000 and 30,000 dimers per cell \rev{in E. coli} during growing and stationary phase, respectively~\cite{Verma2019a,Azam1999a}. IHF binds preferentially to a consensus sequence with high affinity (dissociation constant $K_d \simeq 2$nM) but also non-specifically ($K_d \simeq 2 \mu$M)~\cite{Wang1995ihf} and creates among the sharpest DNA bends in nature, up to 150$^\circ$~\cite{Yoshua2021}. It plays a key role in horizontal gene transfer, integration and excision of phage \ldna~\cite{Laxmikanthan2016} and DNA looping~\cite{Huo2009}. Recent evidence suggest that IHF may also mediate DNA bridging through non-specific, weak interactions \rev{which transiently stabilise distal DNA segments in 3D proximity}~\cite{Yoshua2021}. Additionally, IHF appears to strengthen biofilms by interacting with extracellular DNA~\cite{Devaraj2019}. In light of this evidence, it remains unclear how IHF \rrev{affects DNA entanglements in dense conditions, such as those of the bacterial nucleoid}.

In this Letter we tackle this open question by coupling Molecular Dynamics (MD) simulations and microrheology experiments. Our MD simulations suggest that IHF can speed up the dynamics of long DNA by reducing entanglements. We validate these predictions using microrheology on solutions of entangled \ldna at volume fractions comparable to that of bacterial nucleoid ($\simeq 2\%$). Our results suggest that IHF may act as a ``fluidiser'' by reducing entanglements between DNA molecules and lowering the effective viscosity. By extrapolating our findings to the {\textit E. coli} genome, we argue that at physiological stoichiometries IHF may reduce the effective viscosity of the nucleoid $\sim$200-fold, potentially facilitating genome reorganisation and segregation. 

\begin{figure*}[t!]
    \centering
    \includegraphics[width=0.95\textwidth]{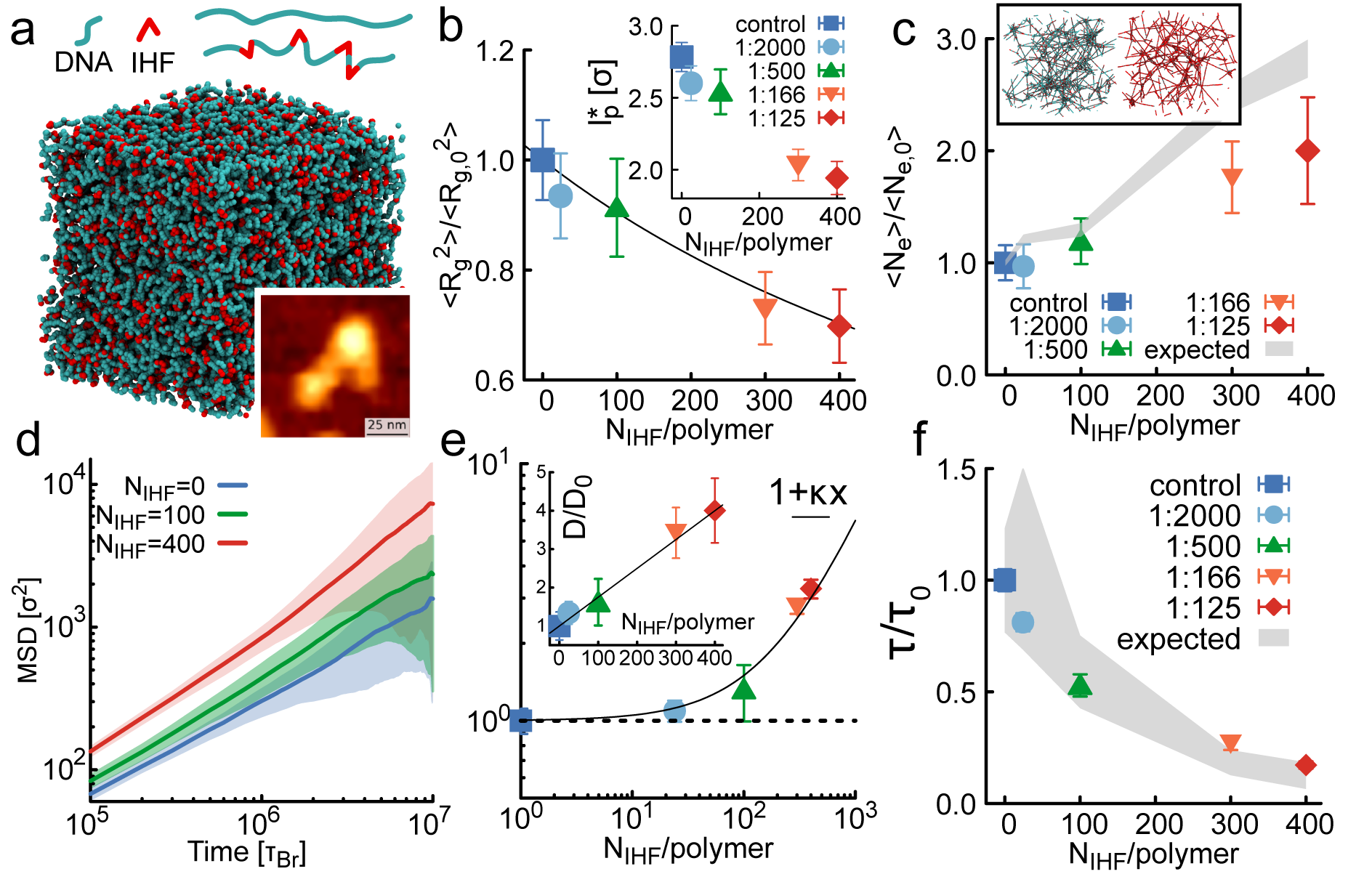}
    \vspace{-0.35cm}
    \caption{\textbf{Molecular Dynamics simulations of kinked semiflexible polymers}. \textbf{a.} Snapshot of the simulation box ($M=50$ chains $N=1,000$ beads long with persistence length $l_p=3 \sigma$ at volume fraction $\rho =0.05$. IHF is modelled as static and stiff harmonic angles forcing \rev{107}$^\circ$ kinks randomly placed along the chains. See also SI Movies. The inset shows an AFM image of a short DNA bound by IHF from Ref.~\cite{Yoshua2021}. \textbf{b.} Normalised squared radius of gyration. (inset) Effective persistence length $l_p^* = 3\langle R_g^2 \rangle/N$. \textbf{c.} Entanglement length from Primitive Path Analysis. Grey shaded area represents predicted $N_e^*(l_p^*)$ with appropriate propagation of errors. (inset) Snapshots from PPA. \textbf{d.} MSD of the centre of mass of the chains. \textbf{e.} Normalised diffusion coefficient. The fitted curve is $1+ \kappa x$ with $\kappa = 0.05$ \rev{(in units of number of IHF in a polymer of 1,000 beads)}. \textbf{f.} Relaxation time $\tau$ defined as $MSD(\tau) \equiv \langle R_g^2 \rangle$. The shaded area represents the values expected using the numerical values of $N_e$ measured in \textbf{c} with appropriate propagation of errors.
    }
     \vspace{-0.5cm}
    \label{fig:panel-ihf-sims}
\end{figure*}

\paragraph*{MD simulations of Entangled DNA with IHF -- } We model solutions of naked \ldna molecules using a variation of the Kremer-Grest model~\cite{Kremer1990} to account for chain stiffness. We simulate $\hbox{M=50}$ coarse-grained bead-spring polymers $N=1,000$ beads long where each bead has size $\sigma = 50$ bp, persistence length $l_p = 3\sigma = 150$ bp and volume fraction $\rho = 0.05$ (see Fig.~\ref{fig:panel-ihf-sims}a). With these choices, each polymer maps to \ldna (48,502 bp) and the expected entanglement length is $N_e \simeq 146$ beads $\simeq 7,300$ bp~\cite{Uchida2008} (see \cite{SM}).
The beads interact via a cut-and-shift Lennard-Jones potential and are connected by FENE springs to avoid chain crossings~\cite{Kremer1990}. Each chain is $N/N_e \simeq 7$ entanglement lengths long. \rev{With these choices, our systems are in the loosely entangled regime~\cite{Morse1998}}. IHF dimers are modelled as permanent stiff harmonic angles constraining triplets of consecutive beads to be bent at 107$^\circ$ (the most frequent angle observed in AFM~\cite{Yoshua2021}) and we neglects unspecific bridging. The simulations are evolved \rev{with implicit solvent (Langevin dynamics)} at $T = 1.0 \epsilon/k_\text{B}$ and timestep $dt=0.01 \tau_{\text{Br}}$ ($\tau_{\text{Br}} = k_\text{B}T/\gamma$ is the Brownian time and $\gamma$ is the friction, set to 1 in LJ units, see SM). 

\begin{figure*}[t!]
    \centering
    \includegraphics[width=0.95\textwidth]{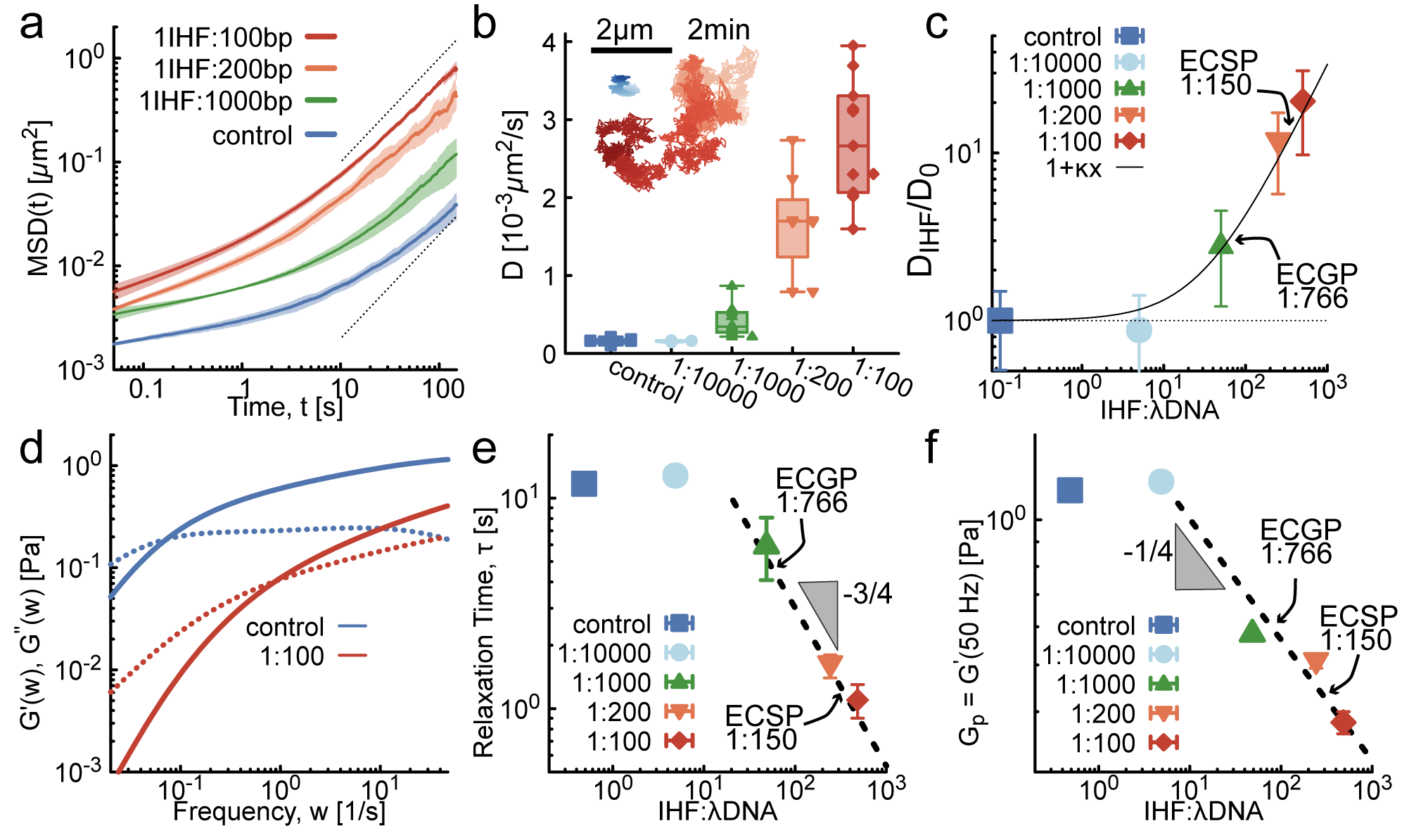}
    \vspace{-0.5 cm}
    \caption{\textbf{Entangled solutions of \ldna are fluidised by IHF}. \textbf{a.} MSDs at different stoichiometries of IHF:DNAbp. The shaded area enveloping the curves is the standard error computed over 10 movies in 3 independent experiments ($>$ 100 tracers total). \textbf{b} Boxplot of the diffusion coefficient $D$ from fitting $MSD = 2Dt$ at large times. (Inset) Representative particle trajectories tracked over 2 minutes. \textbf{c.} Normalised mean diffusion coefficient at increasing concentration of IHF. \textbf{d.} Complex moduli $G^\prime$ (solid) and $G^{\prime \prime}$ (dashed) for the control and for 1 IHF every 100bp. \textbf{e.} Relaxation time $\tau = w_R^{-1}$ where $w_R$ is the crossover frequency at which $G^\prime (w_R) \equiv G^{\prime \prime} (w_R)$. \textbf{f.} Elastic plateau obtained from the value of $G^\prime$ at 50 Hz. We have indicated the two biologically relevant stoichiometries in {\it E. coli} growing and stationary phases as ``ECGP'' and ``ECSP''.
    }
    \vspace{-0.5 cm}
    \label{fig:panel-ihf-conc}
\end{figure*}

To model different IHF stoichiometries, we vary the number of kinks along the chains, let the systems equilibrate, and then perform a production run where we measure the properties and dynamics of the chains. The kinks are placed at random, mimicking non-specific binding. We choose to explore a range of stoichiometries that is physiologically relevant and experimentally feasible {\it in vitro}, i.e. 6,000 and 30,000 IHF dimers in growing and stationary phase~\cite{Azam1999a}, correspond to 1 IHF dimer every 800 and 150 bp within a 4.6 Mbp-long {\it E. coli} genome.

First, we observe that the more the kinks, the smaller the  gyration radius of the chains $\langle R_g^2 \rangle \equiv \langle 1/N \sum_i^N \left[ \bm{r}_i - \bm{r}_{CM} \right]^2 \rangle$ (Fig.~\ref{fig:panel-ihf-sims}b). Due to the self-avoiding interactions being screened in dense solutions~\cite{Doi1988}, we estimate the size of the chain as $R_g = l_p \sqrt{N/3l_p}$, where $l_p$ is the persistence length. In analogy with the case of freely kinked worm-like chains~\cite{Wiggins2005} (albeit here we set the kink to a specific angle rather than leaving a fully flexible joint as in Ref.~\cite{Wiggins2005}) we can renormalise the persistence length to an effective $l_p^*(N_{\rm IHF})$ that depends on the number of kinks introduced in the chains, $N_{\rm IHF}$, and compute it as $l_p^* = 3 \langle R_g^2 \rangle/N$. As shown in Fig.~\ref{fig:panel-ihf-sims}b(inset), the effective persistence length decreases from $l_p = 3\sigma \simeq 150$ bp to around $l_p=1.8 \sigma \simeq 90$ bp when we add 1 IHF every 2.5 beads (or 125 bp). Given that we work at fixed polymer concentration, we use $l_p^*$ to estimate the IHF-dependent entanglement length $N^*_e$ as~\cite{Uchida2008} 
\begin{equation}
N_e^* = l_\text{K}^*\left[ \left( c_\xi \rho_\text{K}^* l^{*3}_\text{K} \right)^{-2/5} + \left( c_\xi \rho_\text{K}^* l^{*3}_\text{K} \right)^{-2} \right]\, ,
\label{eq:uchida}
\end{equation}
where $c_\xi=0.06$, $l^*_\text{K}=2 l^*_p$ is the Kuhn length and $\rho_\text{K} = N M/(l_\text{K} L^3)$ is the number density of Kuhn segments. The grey shaded area in Fig.~\ref{fig:panel-ihf-sims}c shows the expected increase in entanglement length corresponding to the decrease in $l_p^*$ predicted by Eq.~\eqref{eq:uchida}. The actual entanglement length, measured directly via primitive path analysis (PPA)~\cite{Everaers2004} (see SM), is shown as symbols. The actual increase in $N_e$ is more moderate than the prediction yet we still observe a $\sim2$-fold increase, in turn halving the number of entanglements per chain, $N/N_e$. 

To study the dynamics, we compute the mean squared displacement (MSD) of the centre of mass of the chains $g_3(t) = \langle \left[ \bm{r}_{CM}(t+t_0) - \bm{r}_{CM}(t_0) \right]^2\rangle$, where the average is performed over chains and $t_0$. The more the kinks, the faster the dynamics (Fig.~\ref{fig:panel-ihf-sims}d) and the larger the diffusion coefficient $D = \lim_{t \to \infty} MSD/6t$. In the SI movies, one can also visually appreciate these faster dynamics. 

We compute the relaxation time $\tau$ as the time at which a polymer has diffused its own size, i.e. $g_3(\tau) \equiv \langle R_g^2 \rangle$ and we find a scaling compatible with reptation, i.e. $\tau/\tau_{0} \sim (N/N_{e})^3 (N_{e,0}/N)^3 \sim (N_{e,0}/N_e)^3$ (Fig.~\ref{fig:panel-ihf-sims}f), where we used the $N_e$ in Fig.~\ref{fig:panel-ihf-sims}c. Thus, our simulations suggest that IHF-induced kinks drive an effective increase in DNA flexibility which in turn increases the entanglement length (as per Eq.~\eqref{eq:uchida}), reducing the number of entanglements per chain and speeding up the dynamics.

\paragraph*{Microrheology of Entangled DNA with IHF -- } To experimentally validate our predictions, we perform microrheology~\cite{Mason1995,Zhu2008} on entangled \ldna (NEB, 48.5 kbp) at \hbox{1.5 mg/ml}, corresponding to a volume fraction of $1 - 4\%$ for an effective DNA diameter $d=5-10$ nm~\cite{Rybenkov1993}, valid at low salt and with $h=0.34$ nm as the height of one basepair. This is similar to the volume fraction expected in {\textit E. coli} nucleoid. For a 4.6 Mbp genome and $V_{nucleoid} = \pi (0.5 \mu$m$)^2 (2 \mu$m$)\simeq 1.5 \mu$m$^3$ we obtain $\phi \simeq 2\%$ for a $d=5$ nm DNA diameter. Samples are made by mixing 9 $\mu$l of 1.5 mg/ml \ldna (stored in TE buffer) with 1 $\mu$l of native IHF dimers at different concentrations (stored in a 25 mM Tris pH 7.5, 550 mM KCl, 40\% glycerol solution). We track the diffusion of 500 nm tracers spiked in the fluids, and extract their mean squared displacements $\langle \Delta r^2(t) \rangle$ (we have checked that larger bead sizes yield the same results, see SM). In Figs.~\ref{fig:panel-ihf-conc}a,b we show that as little as $50$ IHF dimers per $\lambda$DNA, or $1$ IHF every 1,000 bp (comparable to 1:800 expected in growing phase), can significantly speed up the dynamics. Adding as much $1$ IHF every 100 bp, speeds up the diffusion of the beads $\sim 20$-fold. One can also visually appreciate this speed up from representative trajectories shown in Fig.~\ref{fig:panel-ihf-conc}b(inset). Pleasingly, the normalised diffusion coefficient $D_{\rm IHF}/D_0$ follows the same trend as seen in simulations, i.e. $D(N_{\rm IHF}) = D_0 (1 + \kappa N_{\rm IHF})$ with a linear increase at large, yet physiological, stoichiometries (compare Figs.~\ref{fig:panel-ihf-conc}c and ~\ref{fig:panel-ihf-sims}e)~\cite{SM} %\footnote{We ran controls with different bead sizes, chamber cleaning protocols and post-processing algorithms. We also checked that the adsoprtion of DNA to the glass is not affecting the bulk DNA concentration (see SM for more information).}. 

To further characterise the viscoelastic properties of the system, we use the generalised Stokes-Einstein relation (GSER) to compute the complex stress modulus~\cite{Mason2000} (see SM). The control sample (pure solution of \ldna at $1.5$ mg/ml) displays a pronounced viscoelasticity, with a relaxation time $\tau \simeq 10$ seconds and a high-frequency elastic plateau $G_p \simeq 1$ Pa, in agreement with the values previously obtained via microrheology~\cite{Zhu2008,Teixeira2007} and bulk rheology~\cite{Banik2021} on similar samples (Fig.~\ref{fig:panel-ihf-conc}d). Introducing IHF at physiological stoichiometries significantly affects the rheology of the solution by both decreasing the relaxation timescale, which becomes $\tau \simeq 1$ second at 1:100 IHF:DNAbp (Fig.~\ref{fig:panel-ihf-conc}d,e), and decreasing the elastic plateau to $G_p \simeq 0.3$ Pa (Fig.~\ref{fig:panel-ihf-conc}f).

The elastic plateau $G_p$ is related to the number of entanglements as~\cite{Doi1988,Teixeira2007} $Z = L/L_e = 5 M G_p/(4 \rho N_A k_\text{B} T)$, with $\rho=1.5$ mg/ml and $M= 48,502 \times 650$ g/mol the molecular weight of $\lambda$DNA. We measure $G_p$ as the value of $G^\prime$ at the largest frequency (50 Hz) sampled in this work. \rrev{[Considering the value of $G^\prime$ at the crossover frequency yields the same scaling as $G{^\prime} (50 Hz)$ (see SM).]} For our control, \ldna at 1.5 mg/ml, we find $G_p = 1.23$ Pa yielding $Z \simeq 13$ or $L_{e,0} \simeq 3,700$ bp $\simeq 1.2$ $\mu$m, in line with the one estimated for eukaryotic genomes~\cite{Rosa2008}.

On the other hand, by introducing IHF at 1:100 DNA bp we find that the elastic plateau yields a significantly larger entanglement length $L_{e} \simeq 15,300$ bp $\simeq 5.2$ $\mu$m, corresponding to $Z \simeq 3.1$. We highlight that while the diffusion coefficient of the beads and the viscous and elastic moduli depend on the length of the polymers in solution, the entanglement length $L_e$ does not, and it only depends on polymer concentration and stiffness~\cite{Everaers2004,Uchida2008}. Thus, we can extrapolate our results to infer the level of entanglement in {\textit E. coli} if no NAP or other packaging protein is present as $Z_0 \simeq L_{\rm genome}/L_{e,0} \simeq 1,200$. This implies that the expected relaxation timescale in absence of NAPs should be $\sim \tau_0 Z_0^3 \simeq 55$ years, considering a microscopic disentanglement time of order $\tau_0 = 1$ second (a typical relaxation time for solutions of marginally entangled DNA solutions with $Z \simeq 1$~\cite{Zhu2008}). It is thus clear that the bacterial nucleoid would not be able to undergo segregation unaided by NAPs and other organising proteins. Note that in Fig.~\ref{fig:panel-ihf-conc}c,e,f we have indicated the two biologically relevant stoichiometries in {\textit E. coli} growing and stationary phases as ``ECGP'' and ``ECSP''.

\begin{figure}[t!]
    \centering
    \includegraphics[width=0.51\textwidth]{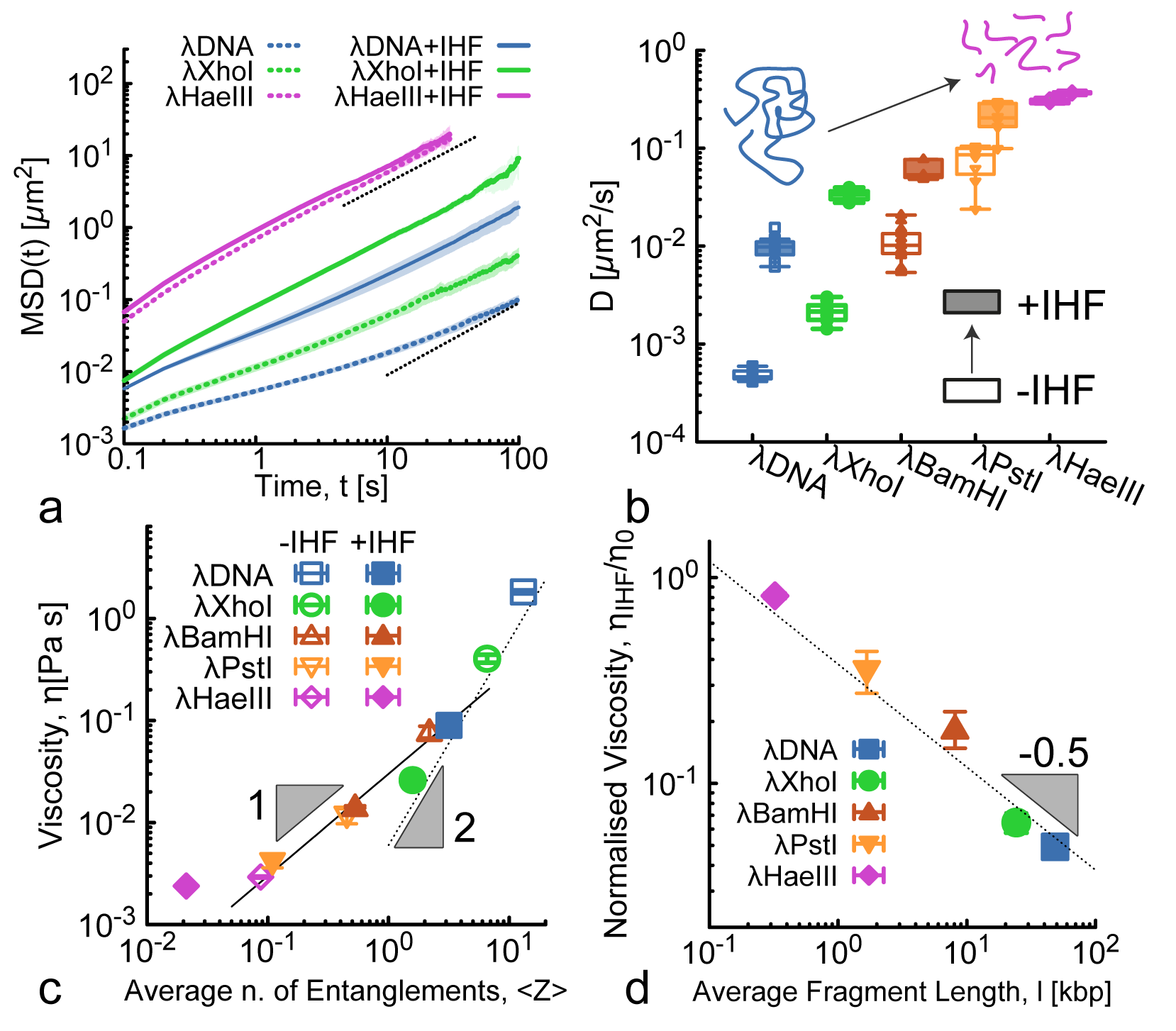}
    \vspace{-0.8cm}
    \caption{ \textbf{Effect of Substrate length on IHF fluidification}. \textbf{a.} MSDs of passive tracers in dense solutions of \ldna pre-digested with different restriction enzymes and before/after addition of 1 IHF every 80 bp. \textbf{b.} Diffusion coefficients of the tracer particles extracted from the large time behaviour as $MSD = 2 D t$. \textbf{c.} Viscosity $\eta$ as a function of average number of entanglements per chain $\langle Z \rangle$. \textbf{d.} Normalised viscosity after/before adding IHF:80bp $\eta_{\rm IHF}/\eta_0$ plotted against average fragment length. In this figure $\lambda$DNA = 0 cuts, $l = 48.5$ kbp; $\lambda$XhoI = 1 cut, $l =24.2$ kbp; $\lambda$BamHI = 5 cuts,  $l =8$ kbp; $\lambda$PstI = 28 cuts, $l=1.7$ kbp; $\lambda$HaeIII = 149 cuts, $l=323$ bp. $l$ is the average fragment length. 
    }
    \label{fig:panel-ihf-dna-length}
    \vspace{-0.6cm}
\end{figure}

In order to use our results to \rrev{obtain insights into} the impact of IHF on the viscoelasticity of the nucleoid {\it in vivo} we address the role of substrate length on the action of IHF. We expect that short, unentangled DNA should be insensitive to the addition of IHF, while longer and deeply entangled DNA should be more affected. To test this hypothesis we perform microrheology on dense solutions (1 mg/ml) of DNA fragments with different lengths but identical overall sequence composition. The samples are obtained by digestion of \ldna via XhoI, BamHI, PstI and HaeIII, restriction enzymes that cut \ldna into 2, 6, 29 and 150 fragments, respectively. As expected, we observe that adding 1:80bp IHF to HaeIII-cut \ldna (referred to as $\lambda$HaeIII) does not affect the MSD of the tracer beads (Fig.~\ref{fig:panel-ihf-dna-length}a). On the contrary, we observe a $\sim$20-fold speed up when IHF is introduced in full length \ldna (Fig.~\ref{fig:panel-ihf-dna-length}a,b)~\footnote{\rrev{Note that we consider a lagtime window of at least 20 seconds in which the exponent of the MSD is compatible with 1 (e.g. 80-100 sec). We then constrain the fit to a linear function $f(t) = 2Dt$.}}. Using Stokes-Einstein, we can compute the viscosity of the samples as $\eta = k_\text{B}T/(3 \pi a D)$ and, by rescaling the average fragment length $l$ by the entanglement length with and without IHF ($L_e=15,300$ and $L_{e,0}=3,700$, respectively), the values of viscosity collapse onto a master curve scaling with the average number of entanglements, $\langle Z \rangle$, as $\eta \sim \langle Z \rangle^{\delta}$ (Fig.~\ref{fig:panel-ihf-dna-length}c). 
\rev{The exponent $\delta=1$ observed at small $\langle Z \rangle$ is expected for Rouse unentangled polymer solutions~\cite{Doi1988}. For $\langle Z \rangle \gtrsim 1$, our data displays a steeper scaling with $\delta=2$. This exponent may be due to the facts that (i) we are in a crossover region to fully reptative systems ($\delta = 3$) and (ii) our systems are polydisperse~\cite{Boudara2019}, as they are generated by cutting \ldna with restriction enzymes~\footnote{We note that accounting for the length-dependence of $L_e$ (becoming shorter at smaller $L$), the scaling may become steeper and closer to the one expected in reptation.}.} 

Intriguingly, by plotting the ratio of the viscosity measured after and before IHF, $\eta_{\rm IHF}/\eta_0$, we observe that the speed up scales with the average length of the DNA fragments as $\eta_{\rm IHF}/\eta_0 \sim l^{-0.5}$ (Fig.~\ref{fig:panel-ihf-dna-length}d). To understand this we have performed MD simulations of entangled polymers of different length (see SM). We found that in the regime investigated in this work the entanglement length $N_e$ has a dependence on the polymer length $N$. More specifically, by adding IHF the entanglement length increases, and the system thus needs longer chains to enter the fully entangled regime. This yields an effective scaling $N_{e,\text{IHF}}/N_{e,0} \sim N^{1/4}$ (see SM). Since the viscosity can be estimated as $\eta = G_e \tau =  G_e \tau_e (N/N_e)^3 \sim N_e^{-2}$, this implies that the ratio, $\eta_{\text{IHF}}/\eta_0 \sim (N_{e,0}/N_{e,\text{IHF}})^{2} \sim N^{-0.5}$ in line with Fig.~\ref{fig:panel-ihf-dna-length}d.

By extrapolating this result to 4.6 Mbp long genomic DNA with about 1 IHF every 100 bp, we expect a reduction in viscosity $\eta_{\rm IHF}/\eta_0 \leq 0.01$, suggesting an effective fludification of {\it E. coli} nucleoid viscosity of about 2 orders of magnitude with respect to the case without IHF. The contribution of other NAPs, transcription factors and genome topology (e.g., supercoiling~\cite{smrek2021topological}) \rrev{will likely affect this estimation and we hope to shed light into these other factors in future works.} 

\paragraph{Conclusions -- } 

In spite of the wealth of single-molecule evidence on how NAPs mechanically interact with short, dilute DNA, the problem of how they regulate entanglements in dense and entangled DNA solutions is poorly understood. We shed light into this problem by performing MD simulations and microrheology on dense \ldna solutions in presence of an abundant NAP called Integration Host Factor (IHF). The key discovery of this work is that IHF acts as a ``fluidiser'' as it reduces the effective viscosity of entangled \ldna by 20-fold at physiological DNA concentrations and IHF:DNA stoichiometries (Figs.~\ref{fig:panel-ihf-sims}-\ref{fig:panel-ihf-conc}). Notably, we measure a quantitatively similar effect by measuring the zero-shear viscosity of DNA solutions via bulk rheology (see SM, Fig.~S7). This fluidification is DNA-length-dependent and we estimate (Fig.~\ref{fig:panel-ihf-dna-length}) that it may shorten the relaxation time of the 4.6 Mbp-long {\textit E. coli} genome by more than $100$-fold. In the future we aim to study systems made of longer, supercoiled DNA and other NAPs such as HNS. We hope that our {\it in vitro} predictions will be tested {\it in vivo} by tracking chromosomal loci in live cells depleted of certain NAPs~\cite{Javer2013}. 

\begin{acknowledgments}
\paragraph{Acknowledgements -- }
DM acknowledges the Royal Society and the European Research Council (grant agreement No 947918, TAP) for funding. JH, ML are supported by Leverhulme Trust (RPG-2017-340) and BBSRC (BB/P000746/1), AN, ML by EPSRC (EP/N027639/1).\\
\end{acknowledgments}
\vspace{ -0.5 cm}

\bibliographystyle{apsrev4-1}
\vspace{ -0.5 cm}
\bibliography{library}

\end{document}

%%%
\title{Fluidification of entanglements by a DNA bending protein: \\ Supplementary Information} 
%%%

%%%
\author{Yair A.~G. Fosado}
\thanks{joint first author}
\affiliation{School of Physics and Astronomy, University of Edinburgh, Peter Guthrie Tait Road, Edinburgh, EH9 3FD, UK}
\author{Jamieson Howard}
\thanks{joint first author}
\affiliation{School of Physics, Engineering and Technology, University of York, York, YO10 5DD, UK}
\author{Simon Weir}
\affiliation{School of Physics and Astronomy, University of Edinburgh, Peter Guthrie Tait Road, Edinburgh, EH9 3FD, UK}
\author{Agnes Noy}
\affiliation{School of Physics, Engineering and Technology, University of York, York, YO10 5DD, UK}
\author{Mark C Leake}
\thanks{co-corresponding author, mark.leake@york.ac.uk}
\affiliation{School of Physics, Engineering and Technology, University of York, York, YO10 5DD, UK}
\affiliation{Department of Biology, University of York, York, YO10 5DD, UK}
\author{Davide Michieletto}
\thanks{corresponding author, davide.michieletto@ed.ac.uk}
\affiliation{School of Physics and Astronomy, University of Edinburgh, Peter Guthrie Tait Road, Edinburgh, EH9 3FD, UK}
\affiliation{MRC Human Genetics Unit, Institute of Genetics and Cancer, University of Edinburgh, Edinburgh EH4 2XU, UK}

\maketitle

\section{Model and Methods}
We model a $\lambda$ DNA molecule 48502 base-pairs (bp) long as a bead-spring polymer made of $1000$ beads. Each bead has a diameter $\sigma=17$ nm (or 50 bp) modelled via a truncated and shifted Lennard-Jones repulsion
\begin{equation}
    U_{\text{LJ}}(r) = 4 \epsilon \left[ (\sigma/r)^{12} - (\sigma/r)^6 + 1/4 \right],
\end{equation}
for $r<r_c=2^{1/6}\sigma$ and 0 otherwise. Here $r$ represents the distance between beads and $\epsilon=1.0$ parametrises the strength of the potential. Consecutive beads are connected through a permanent Finite Extensible Non-linear Elastic (FENE) bond
\begin{equation}
    U_{\text{FENE}}(r) = - 0.5 K R_0^2 \log{\left[1 - \left( r/R_0 \right)^2\right]} 
\end{equation}
with $K=30 \epsilon/\sigma^2$ and $R_0=1.6\sigma$.
The bending stiffness of the polymer is controlled by a Kratky–Porod interaction 
\begin{equation}
    U_{b}(r) = \dfrac{k_\text{B}T l_p}{\sigma} (1 + \cos{\theta}),
\label{eq:Ub}
\end{equation}
which regulates the angle ($\theta$) defined by the two tangent vectors connecting three consecutive beads along the polymer. Here, $l_{p}=3\sigma=150$ bp is the bending persistence length of DNA. Finally, we account for the effective interaction between IHF and DNA through a harmonic potential forcing a sharp angle $\theta_{0}=107^{\circ}$ (or kink) along the polymer,
\begin{equation}
    U_{\text{IHF}}(r) = K_{\theta} (\theta - \theta_{0})^{2},
\end{equation}
where a high value of $K_{\theta}=100$ $k_{B}T$ ensures small fluctuations about the equilibrium angle $\theta_{0}$. It is worth noting here that by modelling IHF in this way, the minimum region of DNA affected by a molecule of IHF is $3$ beads (the ones necessary to define an IHF angle). 
%This is equivalent to 150 bp and therefore, much larger than the real size of IHF ($\sim$10 bp). We make this choice for computational efficiency, a more accurate representation on the size of IHF ($3\sigma = 10$ bp) would require a polymer made of ~14500 beads representing a $\lambda-$DNA molecule. However, we have checked that changes in the resolution of our system give the same qualitative results.

Unless otherwise stated, solutions of naked DNA were simulated by placing $M=50$ of these coarse-grained bead-spring (Kremer-Grest) polymers in a cubic box of size $L=80.6$ $\sigma$ with periodic boundary conditions. The volume fraction of the system is $\rho = \frac{\pi M N \sigma^{3}}{6 L^{3}} =0.05$. According to Eq.~(1) of the main text, for this choice of parameters the expected entanglement length is $N_{e}=146\; \sigma$. Later we will show that this result is in agreement with the Primitive Path Analysis computed from simulations.  Note that each chain is $N/Ne \sim 7$ entanglement lengths, thus placing the system in the entangled regime.

To explore the effect of IHF on on the static and dynamic properties of the entangled polymers we systematically vary the number ($N_{\text{IHF}}$) of IHF dimers in the system. These kinks are placed randomly, mimicking a non-specific binding of IHF along the contour length of any of the polymers. The system was evolved using a velocity-Verlet algorithm, assuming implicit solvent (Langevin dynamics) at temperature $T = 1.0 \epsilon/k_\text{B}$, timestep $dt=0.01 \tau_{\text{Br}}$, where $\tau_{\text{Br}} = k_\text{B}T/\gamma$ is the Brownian time and $\gamma$ is the friction, set to 1 in LJ units.

\begin{figure}[t]
\centering
\includegraphics[width=\columnwidth]{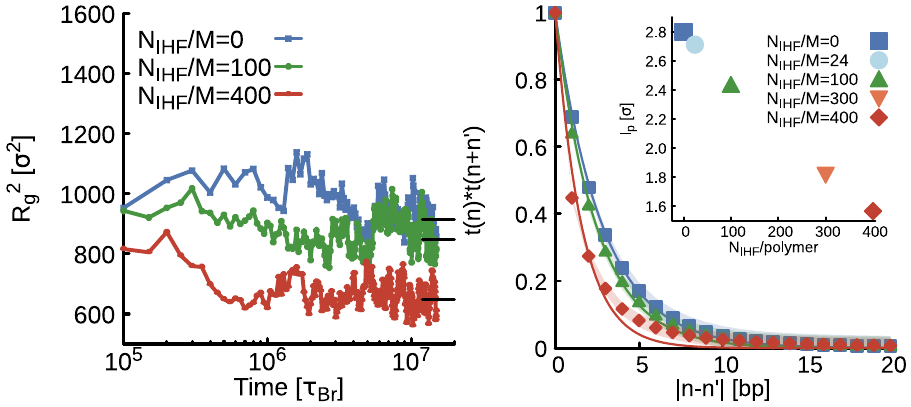}
\caption{\textbf{Statistics of kinked simulated polymers}. (Left) time evolution of the squared radius of gyration $R_g^2$. (Right) tangent-tangent correlation. Results shown here are computed from simulations of a system with $M=50$, $N=1000$, $l_{p}=3\;\sigma$ and $\rho=0.05$. Different colours represent results for different number of IHF interactions per polymer. In left panel, black lines represent the average $\langle R_{g}^2 \rangle$ computed at long times $t>1.2~\times10^{7}$ $\tau_{\text{Br}}$. In right panel points are data from simulations and lines represent a fit using Eq.~\ref{eq.ttc}. The persistence length obtained from this fit is shown in the inset as function of the IHF concentration.}
\label{fig:rg_msd}
\end{figure}

\begin{table*}[t]
\centering
\begin{tabular}{|c|cccc|c|cc|c|}
\toprule
$N_{\text{IHF}}/M$ & $\langle \tau \rangle$ & $\langle R_{g}^2 \rangle$  & $\langle l^{*}_{p} \rangle$ & $\langle N_{e} \rangle$  &$D$    & $\langle N_{e} \rangle$ &  $\langle l^{*}_{p} \rangle$ &  $\langle l^{*}_{p} \rangle$  \\
        &  [$10^{6}\tau_{\text{Br}}$]  & [$\sigma^2$] & [$\sigma$] & [$\sigma$] &[$10^{-5}\sigma^{2}/\tau_{\text{Br}}$] & [$\sigma$]  & [$\sigma$] & [$\sigma$] \\
\midrule
\midrule
  0    & $3.76\pm0.19$ & $913\pm39$ & $2.74\pm0.12$ & $196\pm6$  & $3.16\pm0.58$ & $162\pm13$ & $2.93\pm0.23$ &2.80\\ 
 24    & $3.27\pm0.17$ & $867\pm40$ & $2.60\pm0.12$ & $227\pm8$  & $4.25\pm0.16$ & $157\pm20$ & $2.96\pm0.37$ &2.71\\
 100   & $2.22\pm0.15$ & $847\pm52$ & $2.54\pm0.16$ & $243\pm11$ & $5.12\pm0.98$ & $194\pm18$ & $2.76\pm0.25$ &2.43\\
 300   & $0.98\pm0.06$ & $677\pm36$ & $2.03\pm0.11$ & $464\pm24$ & $10.8\pm0.11$ & $286\pm29$ & $2.40\pm0.25$ &1.82\\
 400   & $0.73\pm0.05$ & $647\pm38$ & $1.94\pm0.11$ & $530\pm32$ & $12.7\pm0.35$ & $325\pm52$ & $2.30\pm0.37$ &1.57\\
\bottomrule
\end{tabular}
\caption{Summary of results from simulations of a system with $M=50$, $N=1000$, $lp=3\;\sigma$, $\rho=0.05$. Columns from left to right show: (1) The number of of IHF kinks per polymer, (2) the relaxation time obtained from the intersection of $R_{g}^2(t)$ and $MSD(t)$, (3) the radius of gyration obtained at long times, (4) the persistence length obtained from Eq.~\ref{lpfromRg} using $R_{g}$, (5) the entanglement length obtained from Eq.~1 in the main text by using the value of $l^{*}_{p}$ obtained before, (6) the diffusion coefficient from the fitting of the $MSD$ at long times, (7) the entanglement length obtained from the PPA analysis, (8) the persistence length obtained from Eq.~1 in the main text by using the previous value of $N_{e}$ and (9) the persistence length obtained from the tangent-tangent correlation (see Fig.~\ref{fig:rg_msd}).}
\label{table:rg2msd}
\end{table*}

\subsection{Equilibration}
In simulations of polymer melts it is important to ensure that equilibrium is reached before gathering and analysing the data. Therefore, our initial configuration was obtained after a long equilibration (1.5 $\times10^{7}$ $\tau_{\text{Br}}$) of a system with $N_{\text{IHF}}=0$. Then, we reset the time-step to zero and we performed simulations with a varying number of kinks per polymer such as $N_{\text{IHF}}/M=0,24,100,300,400$. Each of these simulations was run for another 1.5 $\times10^{7}$ $\tau_{\text{Br}}$.

To assess equilibrium, we compute the temporal evolution of the radius of gyration ($R_{g}$) of each polymer in the simulation by using the relation
\begin{equation}
 R_g^2 (t) =  \dfrac{1}{N}\sum_{n=1}^N \left[ \bm{r}_{CM}(t) - \bm{r}_n(t) \right]^2,
\label{eq:rg}
\end{equation} 
where $\bm{r}_n(t)$, $\bm{r}_{CM}(t)$ and $N$ represent the position of the $n$-th bead of a polymer at time $t$, its centre of mass (COM) and its length, respectively. We also compute the mean squared displacement (MSD) of the COM of each polymer as funtion of the lag-time ($t$)
\begin{equation}
\text{MSD}(t) = \langle \lvert \bm{r}_{CM}(t_{0}+t) - \bm{r}_{CM}(t_{0})  \rvert^{2} \rangle.
\label{eq:msd}
\end{equation}
The equilibration time ($\langle \tau \rangle$) is found as the intersection of the plots $R_g^2(t)$ and $ \text{MSD}(t)$ averaged over all molecules. This is the time it takes to the centre of mass of a polymer to move a distance equal to its own radius. We also confirmed that at this stage the radius of gyration fluctuates around a steady state value $\langle R_{g} \rangle$ (see left panel in Fig.~\ref{fig:rg_msd}). This is the value used in the main plots of Figs.~1b,f. As expected, introducing kinks into the system decreases both $R_{g}$ and $\tau$. From these results we concluded that a reasonable time to start measuring the properties and dynamics of the chains was 5 $\times10^{6}\tau_{\text{Br}}$. In other words, we discarded the first 5 $\times10^{6}\tau_{\text{Br}}$ (equilibration) of all our simulations and we performed the analysis with the remaining data.

The radius of gyration found in simulations can be related to the persistence length ($l^{*}_{p}$) for the Worm-Like Chain (WLC) model. In this model the end-to-end distance ($\mathbf{R}$) of a polymer is given by:
\begin{equation}
\langle \mathbf{R}^{2} \rangle = 2Nl^{*}_{p} \left[ 1 - \frac{l^{*}_{p}}{N} \left( 1 - e^{-N/l^{*}_{p}} \right) \right].
\end{equation}
Then, in the limit $l^{*}_{p} \ll N$ we obtain $\langle \mathbf{R}^{2} \rangle = 2Nl^{*}_{p}$. We also know that for an entropically governed polymer that follows a random walk in three dimensions $\langle R_{g}^{2} \rangle=\langle \mathbf{R}^{2} \rangle/6$, from which we find
\begin{equation}
\langle R_{g}^{2} \rangle = Nl^{*}_{p}/3.
\label{lpfromRg}
\end{equation}
This relation is used to produce the plot in the inset of Fig.~1b in the main text. It shows that adding IHF effectively decreases the persistence length of the system, making it more flexible. By using this value of $l^{*}_{p}$ into Eq.~1 of the main text we obtain the entanglement length displayed by the gray curve in Fig.~1c of the main text.
 
The mean squared displacement computed after equilibration is shown in Fig.~1d of the main text. By fitting the MSD at long times with the relation $MSD(t)=6Dt$, we obtained the diffusion coefficient $D$ in Fig~1e of the main text. All the results presented in this section are summarized in Table~\ref{table:rg2msd}. They show that introducing kinks into the system reduces the number of entanglements ($N/N_{e}$) of the polymers, so their mobility increases. Supplementary movies provide a visual and intuitive support of this. In cyan we represent the trajectory of one polymer from our previous simulations with $N_{\text{IHF}}/M=0$. In red, we show the trajectory of the same polymer but from the simulation with $N_{\text{IHF}}/M=400$. Big beads in blue and red represent the position of the COM of each polymer. At $t=0$ the COM of both polymers have the same position, which is represented by the black bead. It is observed that during the course of the simulation the COM of the red chain moves a larger distance and also that the red chain is more compact. The three different movies represent views of the same system from different planes.

\begin{figure*}[t!]
    \centering
    \includegraphics[width=0.95\textwidth]{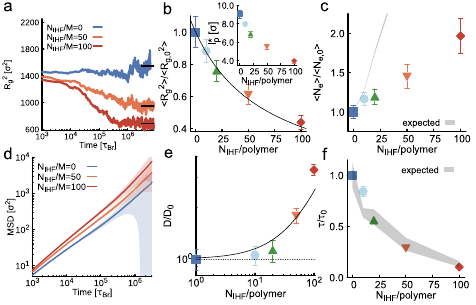}
    \vspace{-0.3cm}
    \caption{\textbf{Molecular Dynamics Simulations of Kinked Semiflexible Polymers}. The system has $M=100$ chains $N=512$ beads long with persistence length $l_p=10 \sigma$ at volume fraction $\rho =0.02$ yielding an entanglement length $N_e \simeq 45$ or $Z=N/N_e \simeq 10$. \textbf{a.} Time evolution of the radius of gyration. \textbf{b.} Normalised squared radius of gyration. (inset) Effective persistence length $l_p^*$. \textbf{c.} Entanglement length from Primitive Path Analysis. Grey shaded area represents predicted $N_e^*(l_p^*)$ with appropriate propagation of errors. \textbf{d.} MSD of the centre of mass of the chains. \textbf{e.} Normalised diffusion coefficient. The fitted curve is $1+ \kappa x$ with $\kappa = 0.015$ (in units of number of IHF in 512 beads). \textbf{f.} Relaxation time $\tau$. The shaded area represents the values expected using the numerical values of $N_e$ measured in \textbf{c} with appropriate propagation of errors.}
     \vspace{-0.5cm}
    \label{fig:panel-ihf-simslp10}
\end{figure*}

\section{Tangent-Tangent correlation}
A different way to obtain the persistence length of a polymer is by measuring the length-scale over which the direction of the chain is no longer correlated with itself, i.e., by computing the tangent–tangent correlator:
\begin{equation}
\langle \mathbf{t}(n) \cdot \mathbf{t}(n') \rangle = e^{- \lvert n-n' \rvert /l^{*}_{p}},
\label{eq.ttc}
\end{equation}
where the tangent at the $n-th$ bead is
\begin{equation}
\mathbf{t}(n)=\frac{\mathbf{r}_{n+1}-\mathbf{r}_{n}}{\lvert \mathbf{r}_{n+1}-\mathbf{r}_{n} \rvert},
\end{equation}
and $\mathbf{r}_{n}$ represents the position of the bead. Results from simulations with different number of kinks are shown in the right panel of Fig.~\ref{fig:rg_msd}.

\begin{table*}[t]
\centering
\begin{tabular}{|c|cccc|c|cc|c|}
\toprule
$N_{\text{IHF}}/M$ & $\langle \tau \rangle$ & $\langle R_{g}^2 \rangle$  & $\langle l^{*}_{p} \rangle$ & $\langle N_{e} \rangle$  &$D$    & $\langle N_{e} \rangle$ &  $\langle l^{*}_{p} \rangle$ &  $\langle l^{*}_{p} \rangle$  \\
        &  [$10^{6}\tau_{\text{Br}}$]  & [$\sigma^2$] & [$\sigma$] & [$\sigma$] &[$10^{-5}\sigma^{2}/\tau_{\text{Br}}$] & [$\sigma$]  & [$\sigma$] & [$\sigma$] \\
\midrule
\midrule
  0   & $2.48\pm0.05$ & $1546\pm73$ & $9.06\pm0.43$ & $52\pm0.4$ & $11.7\pm0.8$ & $86\pm4$   & $14.1\pm0.6$ & 9.73\\ 
 10   & $2.09\pm0.08$ & $1363\pm44$ & $7.98\pm0.26$ & $66\pm0.5$ & $12.4\pm0.7$ & $100\pm3$  & $13.2\pm0.3$ & 8.49\\
 20   & $1.38\pm0.06$ & $1173\pm48$ & $6.87\pm0.28$ & $91\pm1$   & $13.2\pm0.1$ & $102\pm4$  & $13.1\pm0.6$ & 7.67\\
 50   & $0.72\pm0.05$ & $945\pm48$  & $5.54\pm0.28$ & $156\pm3$  & $21.1\pm0.8$ & $124\pm8$  & $12.1\pm0.8$ & 5.90\\
 100  & $0.26\pm0.02$ & $678\pm36$  & $3.98\pm0.21$ & $390\pm10$ & $39.1\pm0.2$ & $169\pm12$ & $10.8\pm0.7$ & 4.18\\
\bottomrule
\end{tabular}
\caption{Summary of results from simulations of a system with $M=100$, $N=512$, $lp=10\;\sigma$, $\rho=0.02$. Columns from left to right show: (1) The number of of IHF kinks per polymer, (2) the relaxation time obtained from the intersection of $R_{g}^2(t)$ and $MSD(t)$, (3) the radius of gyration obtained at long times, (4) the persistence length obtained from Eq.~\ref{lpfromRg} using $R_{g}$, (5) the entanglement length obtained from Eq.~1 in the main text by using the value of $l^{*}_{p}$ obtained before, (6) the diffusion coefficient from the fitting of the $MSD$ at long times, (7) the entanglement length obtained from the PPA analysis, (8) the persistence length obtained from Eq.~1 in the main text by using the previous value of $N_{e}$ and (9) the persistence length obtained from the tangent-tangent correlation.}
\label{table:lp10}
\end{table*}

\section{Primitive Path Analysis}
We performed simulations of the primitive path analysis as described in reference~\cite{SukumaranSathishK2005Itpp}. That is, starting from an equilibrated configuration of our polymer melts, we set the velocity of all particles to zero. In the FENE potential we set $K=100 \epsilon/\sigma$ and $R_{0}=1.2\sigma$, this to prevent chain crossings and ensure the topology conservation. We decreased the timestep from 0.01 $\tau_{\text{Br}}$ to 0.001 $\tau_{\text{Br}}$. Intra-chain excluded volume interactions are disabled while inter-chain ones are retained. The bending persistence length $l_{p}$ is set to zero so bonded beads interact only through a FENE potential with new minimum located at $r=0$. The temperature of the system is lowered to $T=0.001 \epsilon/k_{B}$ and therefore thermal fluctuations are negligible. We used a friction $\gamma=40$ in the first $10^{3}$ time-steps of the simulation to facilitate energy dissipation. After this, $\gamma$ is restarted to its usual value (1) and the system is equilibrated for 5 $\times10^{5}$ timesteps. Under these conditions the chains tend to reduce to the shortest paths between their end-points. In doing so, inter-chain interactions prevent crossings and therefore entanglements remain frustrated. From the final configurations of the chains we compute the entanglement lengths as $\langle N_e \rangle = \langle R_{ee}^2/N b_{pp}^2 \rangle$ where $R_{ee}$ is the end-to-end distance and $b_{pp}$ is the average bond length in the primitive path. The average is performed over all chains in the system and different equilibrated configurations. Results from simulations are shown in Table~\ref{table:rg2msd} and Fig.~1c of the main text. Following to reptation theory, the relaxation time of the system is expected to be proportional to  $\tau/\tau_{0}=(N_{e,0}/N_{e})^{3}$. This prediction is represented by the gray curve in Fig.~1f in the main text, which accounts for errors in the measurements of $\langle N_e \rangle$ via standard error propagation.

\section{Probing the effect of IHF in stiffer polymers}
In the simulations we show in the main text we considered a persistence length of the polymers ($l_p=3\sigma$) that is the same as the angle imposed to mimic IHF (defined by the interaction between 3 consecutive beads). Because of this, we wondered whether there may be some weakening of the kinking imposed by IHF in the simulations. In other words, we expect that semi-flexible polymers with $l_p > 3\sigma$ should be more affected by introducing kinks along the backbone. 

To verify this, we performed simulations and associated analysis on systems of polymers with $l_{p}=10$. More specifically, we consider a solution of DNA with $M=100$ molecules, $N=512$ beads each, in a cubic box of size $L=110 \sigma$ and $\rho=0.02$ for which (according to Refs.~\cite{Everaers2004,Uchida2008}) we expect a similar level of entanglement as the system simulated before.

Results from simulations with a varying number of IHF angles per polymer ($N_{\text{IHF}}/M=0, 10,20,50,100$) are shown in Fig.~\ref{fig:panel-ihf-simslp10}. We stress that at the largest concentration used here ($N_{\text{IHF}}/M=100$), on average, only a fraction $f_{\text{IHF}}=N_{\text{IHF}}/(N-2)M = 20\%$ of the available angle-sites are covered. This is in contrast with the $f_{\text{IHF}}=40\%$ sites covered in the system described in the main text at its largest concentration ($N_{\text{IHF}}/M=400$). Remarkably, the effect of IHF in the new system is more dramatic despite $f_{\text{IHF}}$ being smaller. For instance, Figs.~\ref{fig:panel-ihf-simslp10}a-b show that $R_{g}$ decreases by more than a half when increasing $f_{\text{IHF}}$ from 0\% to 20\%. In comparison, Fig.~1b in the main text shows that $R_{g}$ decreased by less than a half in going from $f_{\text{IHF}}=0$ to 40\%.

In the inset of Fig.~\ref{fig:panel-ihf-simslp10}b we show the persistence length computed from Eq.~\ref{lpfromRg}. However, we recall that this equation is only valid in the random walk approximation, $N \ll l_{p}$. This might be the reason why the entanglement length (Fig.~\ref{fig:panel-ihf-simslp10}c) from the PPA analysis (colour) and the one obtained using Eq.~1 of the main text (gray curve) do not match. The MSD and the diffusion coefficient computed from simulations are shown in Figs.~\ref{fig:panel-ihf-simslp10}d-e. The relaxation time in Fig.~\ref{fig:panel-ihf-simslp10}f is obtained from the intersection of $R_{g}^2(t)$ and $MSD(t)$. 

All the results presented in this section are summarised in Table~\ref{table:lp10}. It is worth mentioning here that in the absence of kinks we expect the persistence length to be $l_{p}=10\sigma$, the one set in the simulations. Interestingly, we obtained different values depending the way in which $l_{p}$ was computed. In the random walk approximation (Eq.~\ref{lpfromRg}) we get $l_{p}=9\sigma$. We attributed this to the fact that the condition $l_{p} \ll N$ is not satisfied. From the tangent-tangent correlation (Eq.~\ref{eq.ttc}) we obtain $l_{p}=9.73\sigma$, close to the expected value. Finally, using the value of $N_{e}$ from the PPA analysis we get $l_{p}=14$, indicating that Eq.~1 of the main text is not a good approximation when $l_{p}$ is about the same order of magnitude as $N$. Importantly, results shown here have the same tendency as before (for the system with $l_{p}=3\sigma$), and qualitatively match our previous findings Fig.~\ref{fig:panel-ihf-simslp10}b.

\begin{figure*}[t!]
    \centering
    \includegraphics[width=0.93\textwidth]{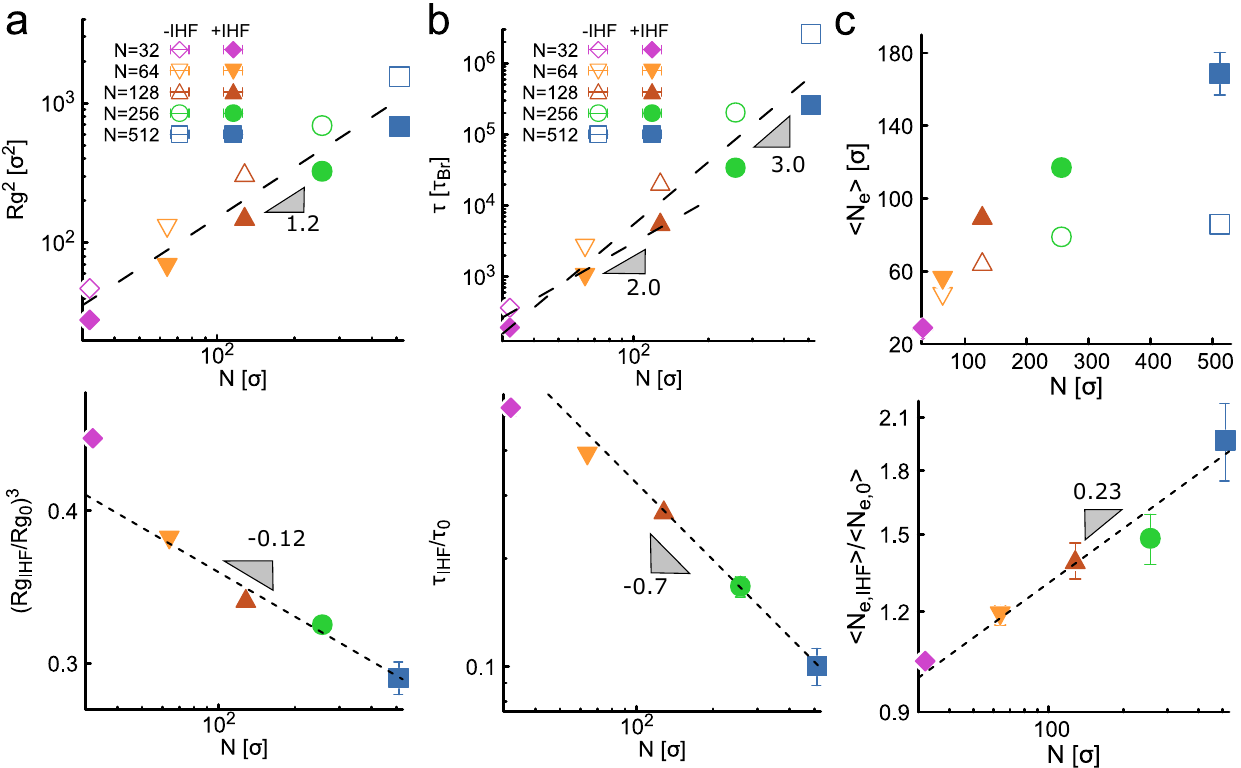}
    \vspace{-0.3cm}
    \caption{\textbf{Molecular Dynamics Simulations of Kinked Semiflexible Polymers of different length at $\rho=0.02$}. Top panels show with empty symbols results from simulations with no IHF kinks (-IHF) and with filled symbols results with a fixed total number of kinks $N_{\text{IHF}}=10000$ (+IHF). Bottom panels show the ratio (data with IHF divided by data with no IHF) from the top panels. \textbf{a.} Squared radius of gyration. \textbf{b.} Relaxation time $\tau$. \textbf{c.} Entanglement length from Primitive Path Analysis. All panels apart from \textbf{c}(top) are presented in Log-Log plots. Dashed lines representing the scaling exponents with polymer size are obtained as a fit to the data in bottom panels while in top panels are a guide to the eye. The relaxation time is expected to scale as $N^{3}$ in entangled systems (reptation theory) and as $N^{2}$ for un-entangled polymers (Rouse theory).}
     \vspace{-0.5cm}
    \label{fig:changeN}
\end{figure*}

\section{Varying the length of the polymers}
In this section we investigate the effect of length on the fluidification of polymer melts. Monodisperse polymers at different lengths are obtained after dividing each of the polymers in the system described above (with $l_{p}=10\sigma$) into $s=1,2,4,8,16$ fragments of the same size $N=512,256, 126,64,32$ and obtaining therefore $M=100, 200, 400, 800, 1600$ molecules. For each of these substrates the volume fraction is constant ($\rho=0.02$) and we run simulations with either, a fixed number of IHF interactions ($N_{\text{IHF}}=10000$) or none ($N_{\text{IHF}}=0$). Figure.~\ref{fig:changeN}a shows that the squared radius of gyration as function of the polymer size behaves as a self-avoiding-walk (SAW) in three-dimensions: $R_{g}^{2} \sim N^{2\nu}$, with $\nu=3/5$. The value of $R_{g}$ is consistently larger in simulations without IHF ($R_{g,0} > R_{g,IHF}$). In presence of IHF, the radius of gyration appears to scale more closely to that of an ideal walk ($\nu = 1/2$). The ratio of the overlapping concentrations measured with and without IHF scales as $c^{*}_{\text{IHF}}/c^{*}_{0}=(R_{g,0}/R_{g,IHF})^{3}$ and according to simulations, this ratio scales as $(R_{g,0}/R_{g,IHF})^{3} \sim N^{-1/8}$ (see bottom panel in Fig.~\ref{fig:changeN}a). We expect that in the large length regime this ratio to approach a constant and length-independent value.

The relaxation time $\tau$ obtained from the intersection of the $MSD(t)$ and $R_{g}(t)$ after equilibration is depicted in Fig.~\ref{fig:changeN}b. Top panel shows that by adding IHF the relaxation of the system is faster ($\tau_{\text{IHF}}<\tau_{0}$) across all lengths. These results can be explained in terms of the reptation model. Consider an entangled chain made of $N$ beads, each of size $b=\sigma$. The movement of the chain is constraint by a confining tube of diameter $a\sim b \sqrt{N_{e}}$ and average length $L=a N/N_{e}$. The diffusion coefficient of such chain is $D_{c}=k_{B}T/N\xi$, with $\xi$ the friction. The time for the chain to diffuse out the confining tube is
\begin{equation}
\tau \sim \frac{L^2}{D_{c}} = \frac{\xi b^{2}}{k_{B}T} N^2_{e} \left( \frac{N}{N_{e}} \right) ^3 = \tau_{e} \left( \frac{N}{N_{e}} \right) ^3,
\label{eq.taurep}
\end{equation}
where $\tau_{e}=\frac{\xi b^{2}}{k_{B}T} N^2_{e}$ is identified with the Rouse time that it takes to an un-entangled chain made by $N_{e}$ monomers of size $b$ to diffuse its own size. 
The ratio of the reptation timescales with and without IHF appears to follow a slightly different scaling and as a consequence of this the ratio $\tau_{\text{IHF}}/\tau_0 \sim N^{-0.7}$. Again, we expect in the very long chain regime this ratio to approach a length-independent value.

Finally, for the entanglement length $N_e$, computed via the PPA described above, we find that its value depends on the chain length (see Fig.~\ref{fig:changeN}c). This is likely due to finite size effects and in the large length limit we observe that the control (no IHF) system appears to converge to a plateau. On the contrary, in presence of IHF this plateau is pushed to longer chains as they are effectively more flexible. In turn, it is necessary to reach longer lengths to be fully in the entangled regime for which the PPA gives more accurate results~\cite{Dietz2022}. We find that in the range of lengths considered here, the ratio $N_{e,IHF}/N_{e,0} \sim N^{0.23}$ indicating that introducing IHF significantly increases the entanglement length in the system.

\begin{figure}[t!]
    \centering
    \includegraphics[width=0.5\textwidth]{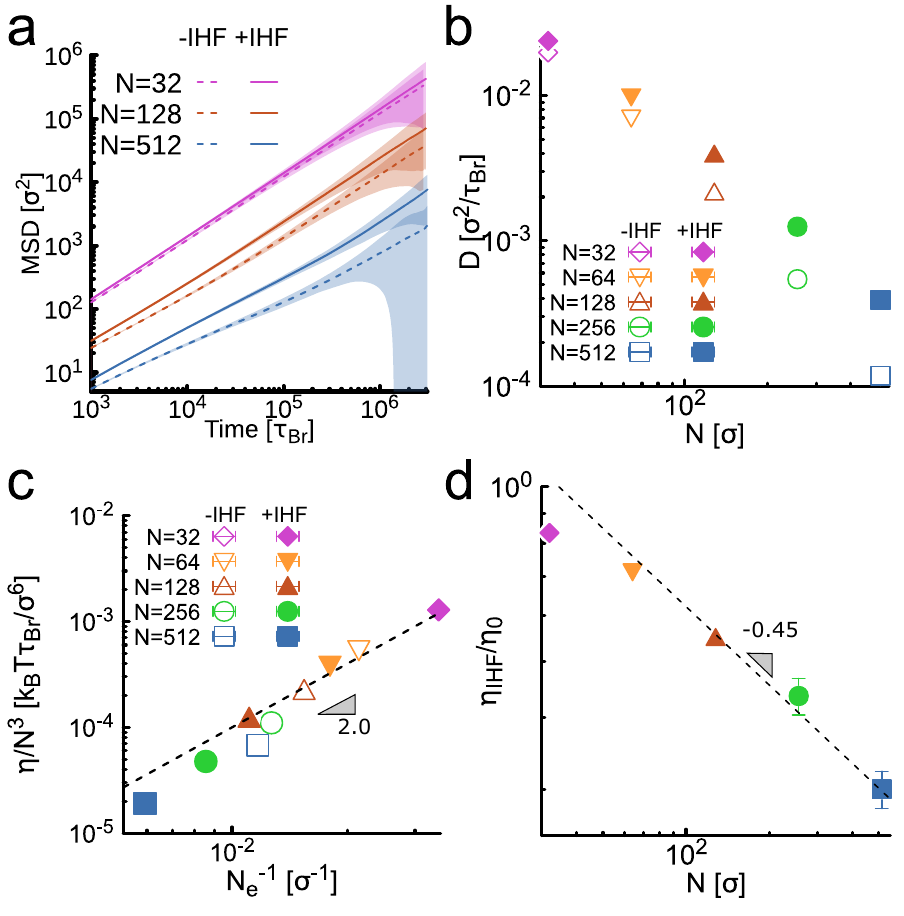}
    \vspace{-0.3cm}
    \caption{\textbf{Simulations of polymers with different lengths}. \textbf{a.} MSD computed from simulations of polymer solutions at the same volume fraction $\rho=0.02$ before and after adding IHF. \textbf{b.} Diffusion coefficient extracted from the MSD at large times using $MSD(t)=6Dt$. \textbf{c} Viscosity computed from $D$ (as $\eta\sim 1/D$) divided by the cubic of polymer length ($N^{3}$) as function of $1/N_{e}$. Data scales as $N_{e}^{-2}$, following the expected relation: $\eta \sim G_{e}\tau_{e}(N/N_{e})^{3}\sim N^{3}N_{e}^{-2}$. \textbf{d.} Normalized viscosity after/before adding IHF.}
     \vspace{-0.5cm}
    \label{fig:DetavsN}
\end{figure}

The mean squared displacement computed from simulations is shown in Fig.~\ref{fig:DetavsN}a and the corresponding diffusion coefficient in Fig.~\ref{fig:DetavsN}b. The mobility of the polymers is in general larger when adding IHF, but we note that, just as in experiments (see Fig.~3b of the main text), the effect of IHF is less marked for smaller chains.

By putting together the results of this section we can write the viscosity as $\eta = G_e \tau =  G_e \tau_e (N/N_e)^3 \sim N^{3}N_e^{-2}$. This scaling is confirmed in Fig.~\ref{fig:DetavsN}c. We then obtain that the ratio  $\eta_{\text{IHF}}/\eta_0 \sim  (N_{e,0}/N_{e,IHF})^{2} \sim N^{-0.5}$ in the regime considered here as observed in both simulations (see Fig.~\ref{fig:DetavsN}d) and experiments (see Fig.~3d in the main text). 

Numerical results obtained from simulations described in this section are depicted in Table.~\ref{table:changingN}.

\begin{table*}[t]
\centering
\begin{tabular}{|cc|cccc|cc|cc|cc|}
\toprule
$N$ & $M$ & $\langle \tau_{0} \rangle$ & $\langle \tau_{\text{IHF}} \rangle$ & $\langle R^{2}_{g,0} \rangle$ & $\langle R^{2}_{g,IHF} \rangle$  & $D_{0}$ & $D_{\text{IHF}}$  & $\langle N_{e,0} \rangle$ & $\langle N_{e,IHF} \rangle$ &  $\langle l_{p,0} \rangle$  &  $\langle l_{p,IHF} \rangle$\\
    &     &  [$10^{3}\tau_{\text{Br}}$]  & [$10^{6}\tau_{\text{Br}}$] & [$\sigma^2$] & [$\sigma^2$] & [$10^{-5}\sigma^{2}/\tau_{\text{Br}}$] & [$10^{-5}\sigma^{2}/\tau_{\text{Br}}$] & [$\sigma$] & [$\sigma$] & [$\sigma$] & [$\sigma$]\\
\midrule
\midrule
32  &1600 & $0.37   \pm0.01$   & $0.19  \pm0.01$  & $47 \pm0.3$ & $28\pm0.2$ & $1973\pm26$ &  $2371\pm19.2$ & $28\pm0.01$ & $29\pm0.15$  & $9.44$ & 3.93 \\ 
64  & 800 & $2.62   \pm0.04$   & $1.02  \pm0.02$  & $129\pm2$   & $67\pm1$   & $704\pm12$  &  $985 \pm 15$  & $47\pm1.2$  & $56\pm0.6$   & $9.45$ & 4.04 \\
128 & 400 & $21.20  \pm0.44$   & $5.76  \pm0.13$  & $313\pm6$   & $152\pm3$  & $214\pm508$ &  $391 \pm10$   & $64\pm2.1$  & $89\pm1.6$   & $9.49$ & 4.13 \\
256 & 200 & $204.14 \pm6.04$   & $34.19 \pm1.25$  & $691\pm18$  & $325\pm11$ & $54\pm1.2$  &  $125 \pm6$    & $81\pm5.7$  & $115\pm0.2$  & $9.51$ & 4.19 \\
512 & 100 & $2586.70\pm155.67$ & $259.71\pm15.29$ & $1546\pm73$ & $678\pm36$ & $11.7\pm0.8$&  $39  \pm0.2$  & $86\pm3.6$  & $169\pm11.7$ & $9.73$ & 4.18 \\
\bottomrule
\end{tabular}
\caption{Summary of results from simulations at fixed volume fraction ($\rho=0.02$) and total number of IHF kinks. Results with $N_{\text{IHF}}=10000$ (or 0) are represented by notations with subindex $_{\text{IHF}}$ (or $_{0}$). The persistence length set in simulation (see Eq.~\ref{eq:Ub}) is $lp=10$. Columns from left to right show: (1) The size of each polymer, (2) the number of molecules in the system, (3-4) the relaxation time obtained from the intersection of $R_{g}^2(t)$ and $MSD(t)$, (5-6) the radius of gyration obtained at long times, (7-8) the diffusion coefficient from the fitting of the $MSD$ at long times, (9-10) the entanglement length obtained from the PPA analysis and (11-12) the persistence length obtained from the tangent-tangent correlation.}
\label{table:changingN}
\end{table*}

\section{Generalised Stokes Einstein Relation}
In this section we discuss in more detail how we perform the conversion from MSD curves of the tracer beads in the system to the stress relaxation modulus $G^*(w)$. This is done via the so-called ``Generalised Stokes Einstein Relation''~\cite{Mason1995,Mason2000}, i.e. in its Laplace formulation
\begin{equation}
    \tilde{G}^*(s) = \dfrac{d k_\text{B}T}{3 \pi a s \langle \Delta \tilde{r}^2(s) \rangle} \, ,
\end{equation}
where $d$ is the dimensionality of the MSD and  $\langle \Delta \tilde{r}^2(s) \rangle$ the Laplace transform of the MSD. To obtain the complex modulus, the numerical data can be fitted by an analyitical function which is then analytically continued into the complex domain as $s \to i \omega$. 
In practice this equation can be computed directly into the Fourier domain and approximated by~\cite{Mason2000}
\begin{equation}
    |G^*(\omega)| = \dfrac{d k_\text{B}T}{3 \pi a i \omega \langle \Delta r^2(1/\omega) \rangle \Gamma[1+\alpha(\omega)])} \, ,
\end{equation}
where 
\begin{equation}
    \alpha(\omega) = \left. \dfrac{d \log{\langle \Delta r^2(t) \rangle}}{d \log t} \right|_{t=1/\omega}
\end{equation}
and obtain the elastic and viscous moduli as 
\begin{align}
    &G^\prime(\omega) = |G^*(\omega)| \cos{ (\pi \alpha(\omega)/2)} \\ 
    &G^{\prime \prime}(\omega) = |G^*(\omega)| \sin{ (\pi \alpha(\omega)/2)} \,
\end{align}
respectively. 

\rev{\section{IHF purification} 
IHF was produced in the E. coli strain BL21AI containing the plasmid pRC188 (a gift from the Chalmers laboratory, the University of Nottingham, UK). The \emph{E. coli} harbouring pRC188 were grown to an OD600 $\sim$ 0.6 in 2L LB + 100 $\mu$g/ml carbenicillin at 37$^\circ$C with shaking at 180 rpm. Induction of IHF overexpression was carried out by the addition of arabinose and IPTG to respective final concentrations  of 0.2\% (w/v) and 1 mM, growth was then allowed to continue for a further 3 hours at 37$^\circ$C with shaking at 180 rpm. The \emph{E. coli} were collected via centrifugation at 3500 x g for 20 minutes at 4$^\circ$C. After discarding the supernatant cells were resuspended in 10 mM Tris pH 7.5, 10\% sucrose (w/v) before being flash frozen in liquid nitrogen and stored at -80$^\circ$C.
Cells overexpressing IHF were thawed on ice and the buffer was adjusted to contain final concentrations of 50 mM Tris pH 8.4, 150 mM KCl, 20 mM EDTA, 10 mM DTT and 0.2 mg/ml lysozyme. The cell suspension was mixed by inversion and kept on ice for 15 minutes before Brij58 was added to a final concentration of 0.1\% (w/v) followed by further mixing by inversion and a 15 minute incubation on ice. The lysed cells were clarified by centrifugation at 4$^\circ$C and 148,000 x g for 60 minutes. The supernatant was collected and polymin P was added to a final concentration of 0.075\% (w/v) in a dropwise fashion whilst stirring at 4$^\circ$C, stirring was continued for 10 mins before centrifugation at 4$^\circ$C and 30,600 x g for 20 minutes. The resulting supernatant was then subjected to a 50\% ammonium sulfate (AmSO4) precipitation followed by an 80\% AmSO4 precipitation. Pellets and supernatant were collected following centrifugation as above with polymin p, IHF was present in the supernatant at 50\% AmSO4 and in the pellet at 80\% AmSO4. AmSO4 precipitated IHF was dissolved in a sufficient volume of buffer A (50 mM Tris-HCl pH 7.5, 2 mM EDTA, 10 mM $\beta$-ME, 10\% glycerol) such that the conductivity of the sample matched that of buffer A + 100 mM KCl. The sample was loaded onto a self-poured  10 mL P-11 phosphocellulose XK-26 column equilibrated with buffer A + 100 mM KCl, washed with 30 column volumes (CV) of buffer A + 100 mM KCl and developed with a 20 CV gradient of 0.1-1 M KCl in buffer A. Fractions containing IHF were identified by their absorbance at 280 nm and with 15\% SDS polyacrylamide gel electrophoresis, these fractions were pooled. Pooled fractions were dialyzed against buffer A + 100 mM NaCl. The sample was then loaded onto a 5 mL HiTrap Heparin column equilibrated with the same buffer, washed with 6 CV of buffer A + 100 mM NaCl and  developed with a 20 CV gradient of 0.1-1 M NaCl in buffer A. Fractions containing IHF were again identified using absorbance at 280 nm  and 15\% SDS polyacrylamide gel electrophoresis and pooled. Pooled fractions were dialysed against buffer B (25 mM Tris.HCl pH 7.5, 550 mM KCl and 40\% glycerol), the dialysed fractions were then aliquoted and flash frozen in liquid nitrogen before storage at -80$^circ$C. Protein concentrations were determined using the Bradford Protein Assay (Bio-Rad).}

\rev{
\section{Sample Preparation}
Solutions of \ldna and IHF were prepared as follows. \ldna was commercially obtained from NEB. Before use, \ldna was always fully thawed at 65$^\circ$C and then left equilibrating on a roller for 1h. The nominal concentration is 0.5 mg/ml and this was checked with a nanodrop. To concentrated \ldna we used ethanol precipitation, i.e. for a volume V, added V/9 of 3M sodium acetate (NaAc), and 2(V+V/9) of 100\% Ethanol. We then placed the mix at -20$^\circ$C overnight and then spun at 17,000g for 1 hour. Pellets of DNA were clearly identifiable at the bottom. We then removed the supernatant, washed the pellet with 70\% ethanol for 3 times. At the end, the pellet was left to dry for a few minutes and resuspended in TE buffer (10 mM Tris, 1mM EDTA, made with DNAse free water) to reach the desired concentration. The final concentration was then checked with a nanodrop. 
The mix of \ldna and IHF were prepared by mixing 9$\mu$l of DNA, 0.5 $\mu$l of beads (of different sizes, see below), and 1 $\mu$l of IHF (pure from stock or dilute in its own buffer) or pure buffer (25 mM Tris pH 7.5, 550 mM KCl, 40\% glycerol) for negative control. }

\begin{figure}[t!]
    \centering
    \includegraphics[width=0.5\textwidth]{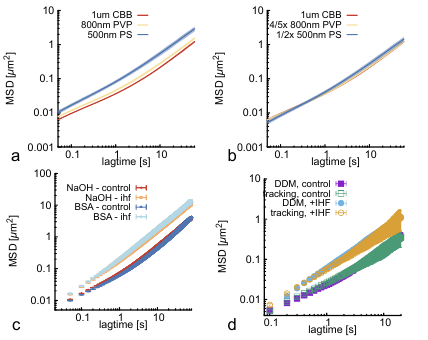}
    \vspace{-0.3cm}
    \caption{\textbf{Controls}. \textbf{a.} MSD of different beads spiked in a solution of \ldna at 0.5 mg/ml. \textbf{b.} Same as \textbf{a} with the MSD scaled by the particle size. \textbf{c} Comparison of the microrheology performed on NaOH cleaned and BSA treated chambers, with and without IHF added to a 0.5mg/ml solution of \ldna. \textbf{d.} Comparison between the MSDs obtained via particle tracking and differential dynamic microscopy (DDM) on the same video. Note that in \textbf{d} we use a single movie while in the other panels (and in the main text) we average across 5-10 movies. One can appreciate that the data points fall on top of each other, within errors.}
     \vspace{-0.5cm}
    \label{fig:controls}
\end{figure}

\rev{
\section{Controls with different slides preparation and bead sizes}
To check that our results are not affected by the choice of tracer size and our sample preparation we have repeated some of our measurements on a 0.5mg/ml solution of \ldna using different bead sizes with different surface treatments: 500 nm polystyrene beads (Sigma), 800 nm Polyvinylpyrrolidone (PVP) coated beads (gifted by Andrew Schofield, Edinburgh) and 1 $\mu$m  Carboxyl Latex Beads (ThermoFisher). As seen in Fig.~\ref{fig:controls}a-b, the beads give MSDs which can be scaled on top of each other within errors. It is known that smaller beads suffer from a depletion layer which tends to underestimate the entanglements in the fluid~\cite{Chen2003}. This effect appears to be small and negligible in these samples most likely due to the fact that the DNA concentration is larger than the one considered in Ref.~\cite{Chen2003}.}

\rev{We have also checked that the concentration of DNA remains the same before and after incubation in the chamber slide. This was done by measuring the 260nm wavelength absorbance with a nanodrop before and after a 5 ul drop of DNA was sandwiched between two slides and left to incubate for 10 minutes (the time typically needed for microrheology measurements). We have not observed any change in DNA concentration after the incubation suggesting that the adsorption of DNA to the glass slide is, if present, negligible with respect to the bulk.}

\rev{In Fig.~\ref{fig:controls}c we also report MSDs of beads on NaOH cleaned and BSA coated slides. The former were cleaned by soaking overnight into NaOH, washed with water and dried with nitrogen before used for sandwiching the DNA samples. The latter were prepared by NaOH cleaning followed by deposition of a BSA (20 mg/ml) drop on the slides and cover slips, left for 20 minutes, followed by washing and drying with nitrogen. As one can appreciate from Fig.~\ref{fig:controls}c, the curves obtained for the samples with and without IHF (prepared as 9 $\mu$l DNA at 0.5mg/ml + 1 $\mu$l of IHF at 245$\mu$M or 1 $\mu$l of IHF storage buffer for the controls) are on top of each other. 
In Fig.~\ref{fig:controls}d we finally report a check done on our postprocessing analysis where we analyse a single movie either with particle tracking or DDM (differential dynamic microscopy)~\cite{Cerbino2008}. One can again appreciate that the two different analysis yield very similar MSDs. Note that the error bars are larger because we are considering only a single movie, while in other plots we average over several (up to 10) movies from independent samples and/or positions in the samples.
}

\begin{figure}[t!]
    \centering
    \includegraphics[width=0.32\textwidth]{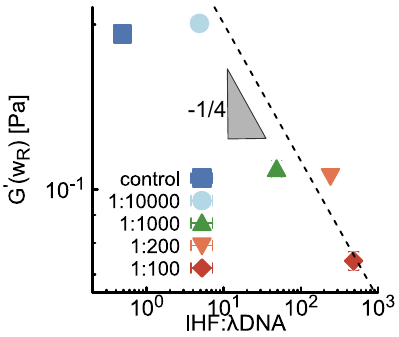}
    \vspace{-0.3cm}
    \caption{Elastic modulus at the cross-over frequency as function of the IHF concentration.}
     \vspace{-0.5cm}
    \label{fig:superposition}
\end{figure}

\rev{
\section{Elasticity scaling}
To determine the IHF-concentration-dependent mechanical properties of our solutions (at 1.5 mg/ml $\lambda$-DNA), here we report the value of $G^\prime$ at the cross-over frequency $w_R$. This provides (at each value of the IHF concentration) a pair of data points ($w_R$, $G^\prime(w_R)$) that can be uniquely and consistently found across all our samples. In Fig.~\ref{fig:superposition} we show the scaling of $G^{\prime}(w_R)$ with the IHF concentration. Importantly, this is the same behaviour as the one reported in Fig.2f of the main text for $G^{\prime}$ at the largest frequency in our experiments (50 Hz). We note that, since $G^{\prime}(w_R)$ represents a low bound value of the elastic plateau ($G_{p}$), a better approximation is: $G_{p}\sim G^{\prime}(\text{50 Hz})$.}

\rev{
\section{Bulk Rheology}
Bulk rheological measurements were  performed using a stress controlled Kinexus 
Pro rheometer by NETZSCH. Samples made of 360 $\mu$l of 0.5mg/ml \ldna and 40 $\mu$l of IHF (at 245 $\mu$M) or IHF buffer were prepared as stated in the main text, minus the beads. All stress tests were performed at 25$^\circ$C using 40 mm stainless steel parallel plate geometries. Samples were carefully pipetted to avoid bubbles formation. Gap height between  plates was maintained at 150-200 $\mu$m and  trimming was performed where needed such that a good fill to the plate edges was achieved. To minimise solvent evaporation a solvent trap was fitted using light mineral oil [Sigma-Aldrich, 330779-1L] in the solvent well. To obtain the shear stress curves, the shear  stress was increased in a step-wise manner,  from 0.01 Pa to $\sim$ 10 Pa, then decreased from $\sim$ 10 Pa to 0.01 Pa. This looping was repeated to ensure that the curve was reproducible and there are no hysteresis effects.}

\begin{figure}[t!]
    \centering
    \includegraphics[width=0.45\textwidth]{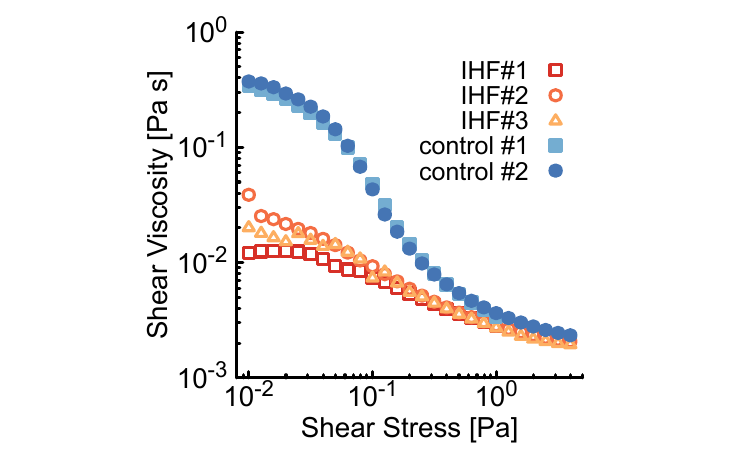}
    \vspace{-0.3cm}
    \caption{\textbf{Bulk rheology}. This figure shows results from bulk rheology stress sweep experiments ranging from 0.01 to 10 Pa. The average zero-shear viscosity for the controls is 15-fold larger than the samples in which IHF is added. }
     \vspace{-0.5cm}
    \label{fig:bulk}
\end{figure}

\rev{The result from independent samples (3 for IHF and 2 controls) stress sweep is shown in Fig.~\ref{fig:bulk}. The zero-shear (low shear stress) values of viscosity for the IHF samples are about 15 times smaller than the viscosity of the controls where IHF is not added (and instead buffer is added). This is in quantitative agreement with the 20-fold decrease observed in the main text. The larger effect seen in the main text is likely due to the fact that the solutions used for microrheology are 3 times more concentrated (1.5mg/ml instead of 0.5mg/ml used for bulk rheology). }

\bibliography{librarySI,library}